\documentclass[sigconf]{acmart}
\setcopyright{none}
\renewcommand\footnotetextcopyrightpermission[1]{}

\AtBeginDocument{%
  }

\begin{document}

\title{Formative Study for AI-assisted Data Visualization}


\author{Rania Saber}
\affiliation{%
  \institution{University of California}
  \city{Riverside}
  \country{USA}}
\email{rsabe004@ucr.edu}

\author{Anna Fariha}
\affiliation{%
  \institution{University of Utah}
  \city{Salt Lake City}
  \country{USA}}
\email{afariha@cs.utah.edu}

\begin{abstract}
This formative study investigates the impact of data quality on AI-assisted data visualizations, focusing on how uncleaned datasets influence the outcomes of these tools. By generating visualizations from datasets with inherent quality issues, the research aims to identify and categorize the specific visualization problems that arise. The study further explores potential methods and tools to address these visualization challenges efficiently and effectively. Although tool development has not yet been undertaken, the findings emphasize enhancing AI visualization tools to handle flawed data better. This research underscores the critical need for more robust, user-friendly solutions that facilitate quicker and easier correction of data and visualization errors, thereby improving the overall reliability and usability of AI-assisted data visualization processes.
\end{abstract}




\received{September 2024}

\maketitle

\section{Introduction}
With the emergence of AI tools such as ChatGPT in today’s data-driven world, the ability to visualize data effectively is crucial for extracting actionable insights from large and complex datasets. AI-assisted data visualization tools have gained prominence due to their capability to automate and enhance the visualization process, making it accessible to a wider audience. However, the effectiveness of these tools is heavily dependent on the quality of the underlying data. Poor data quality, characterized by errors, inconsistencies, and missing values, can lead to inaccurate or misleading visualizations, ultimately compromising decision-making processes.

Despite the growing reliance on AI-driven visualization tools, there is a significant gap in understanding how these tools handle uncleaned datasets. This study addresses this gap by exploring the specific challenges and limitations of AI-assisted visualizations when faced with data quality issues. By generating visualizations using uncleaned datasets, we aim to identify and categorize common visualization problems, providing valuable insights into the weaknesses of current AI tools in processing flawed data.

The primary objective of this research is to investigate methods and tools that could potentially address these visualization issues quickly and effectively. While this study focuses on identifying the problems and proposing potential solutions, it does not yet involve the development of new tools. Instead, it lays the foundation for future work aimed at enhancing the efficiency and usability of AI-assisted data visualization tools.

This research is significant as it highlights the critical need for more robust solutions in the field of AI-assisted data visualization. By understanding the impact of data quality on visualization outcomes, we can begin to design tools that better handle imperfect data, thereby improving the reliability and utility of these visualizations in real-world applications. The findings of this study are expected to inform future efforts to develop faster, more user-friendly tools that facilitate the correction of data and visualization errors, ultimately leading to more accurate and trustworthy visualizations.

\section{Literature Review}
Before conducting experiments, I reviewed several key research papers to provide a solid foundation for my study. These papers offered insights into the relationship between data quality and automated visualizations, as well as the tools and methods used to address common issues. Through this review, I gained a deeper understanding of how poor data quality impacts AI-assisted visualizations and what methods have been proposed to mitigate these problems.

One of the key papers I explored was “Surfacing Visualization Mirages” by \cite{vizmirages}. In this paper, the authors introduced the concept of "visualization mirages"—errors or distortions in visualizations caused by noise, bias, or poor data quality. I was particularly interested in how poor data quality can lead to misleading visualizations and the potential consequences for users who rely on these visualizations for decision-making. The paper's focus on identifying and mitigating these mirages provided valuable insights into the types of data issues that might arise during my own research. Their methods for recognizing and addressing mirages helped shape my approach to analyzing the visualizations generated by AI, particularly when dealing with unclean datasets. The authors emphasized the importance of first identifying the sources of inaccuracies before attempting to resolve them, a principle I adopted in my own experiments.

Building on this foundation, I also reviewed
 “Automated Data Visualization from Natural Language via Large Language Models: An Exploratory Study” by  \cite{Wu_2024}. This study examined the ability of large language models (LLMs) to handle visualization tasks based on natural language input. The authors demonstrated that while LLMs have great potential for automating the creation of data visualizations, they face limitations when dealing with more complex or unclean datasets. This paper was particularly relevant to my study because it highlighted the capabilities and shortcomings of AI models when tasked with generating visualizations, especially in scenarios where data quality is compromised. It helped frame my understanding of how well LLMs handle visualization tasks and what challenges arise when these models are applied to datasets with inherent quality issues.

In addition to these foundational studies, I explored research on tools designed to detect and correct visualization errors. The paper “Linting for Visualization: Towards a Practical Automated Visualization Guidance System” by \cite{Mcnutt2018LintingFV} introduced a linting system that automatically detects and flags errors in visualizations. This system provides real-time feedback to users, enabling them to identify and correct mistakes early in the process. Their research demonstrated the value of automated guidance in improving the accuracy of visualizations, particularly when users are dealing with complex datasets. This paper informed my understanding of how AI-assisted tools can be enhanced by integrating automated error-checking mechanisms.

Similarly, "VisuaLint" by \cite{hopkins} focused on providing real-time annotations to help users detect chart construction errors. Their approach to identifying visualization errors aligns with my research goal of understanding how AI tools can assist users in correcting visualizations based on unclean data. The idea of "sketchy" annotations, as described in this paper, highlighted the potential for AI systems to offer iterative feedback and guidance, which I considered while designing the error correction process in my experiments.

Further insights were gained from "The Data Linter: Lightweight Automated Sanity Checking for ML Data Sets" by \cite{hynes}, which introduced a system for automatically checking the quality of datasets used in machine learning pipelines. This system's lightweight, automated approach to identifying data quality issues underscored the importance of early detection and correction of data problems before visualizations are generated. This paper helped me better understand the role of automated tools in ensuring data integrity and informed my approach to evaluating the impact of unclean data on AI-generated visualizations.

Finally, the work of \cite{vizlinter} in "VizLinter: A Linter and Fixer Framework for Data Visualization" further contributed to my understanding of how automated systems can not only detect errors in visualizations but also suggest fixes. Their framework extended traditional linting by providing users with actionable solutions to correct data quality issues, an approach that I aimed to explore in my own study. The ability of AI tools to offer suggestions for correcting errors, rather than just identifying them, was a key focus of my analysis.

These studies collectively provided a strong foundation for understanding the challenges and opportunities in AI-assisted data visualization, particularly when dealing with poor data quality. While previous research has explored various tools and methods for detecting and correcting visualization errors, there remains a gap in understanding how AI-driven systems cope with unclean datasets. My research builds on these ideas by exploring how data quality issues specifically affect AI-generated visualizations and how these errors manifest when using unclean datasets.

\section{Methodology}
This study was conducted in three phases to explore the impact of data quality on AI-assisted data visualizations. The first phase involved analyzing visualizations created from a clean dataset, while the second phase focused on using an unclean dataset. The third phase consisted of an experimental study where specific data quality issues were systematically injected into clean datasets to observe their impact on visualizations.

\subsection{Phase 1: Clean Dataset Analysis}
A clean dataset, the 911 dataset from Kaggle, was selected for initial analysis. This dataset was chosen due to its well-documented structure and absence of quality issues, making it ideal for baseline comparisons. Ten different visualizations were generated using ChatGPT by providing the first five rows of the dataset and specifying the desired visualization type. Python scripts provided by ChatGPT were executed in Google Colab, and any errors in the visualizations or code were noted. The process of correcting these visualizations was documented, including the number of iterations and steps required to resolve any issues.

\subsection{Phase 2: Unclean Dataset Analysis}
For the second phase, The Metropolitan Museum of Art Open Access dataset from GitHub was used, which included several known quality issues such as missing values, inconsistent information, and possible duplications. The objective was to generate visualizations using ChatGPT, similar to the clean dataset phase, and document the errors and corrections. ChatGPT was allowed three attempts to fix each visualization error. The time taken and the difficulty of resolving each issue were recorded to evaluate the challenges associated with unclean data.

\subsection{Phase 3: Experimental Study with Injected Data Quality Issues
}
In the third phase, ten clean datasets were selected, and eight different data quality issues were systematically injected into these datasets. These issues included missing data, duplicate data, inconsistent data, inconsistent data types, inaccurate data, irrelevant data, data entry errors, and incorrect data formats. Each modified dataset version was used to generate five different types of visualizations: pie charts, word clouds, histograms, heat maps, and scatter plots, totaling fifty visualizations per dataset. Python scripts were developed to automate the injection of data quality issues, such as removing 15\% of values to simulate missing data or randomizing data formats to create inconsistencies.
\\

The visualizations were analyzed to assess how each type of data quality issue affected the outcome. Errors, iterations, and time taken to correct visualizations were meticulously documented, providing insights into the specific challenges posed by different types of flawed data.

\textit{Note on study limitations:}
\\
The experimental study phase was initiated in the seventh week of a ten-week Distributed Research Experiences for Undergraduates (DREU) program. Due to time constraints, not all planned analyses and evaluations of the various injected data quality issues were completed. Specifically, while initial efforts focused on generating and analyzing visualizations with several data quality issues, the comprehensive examination of all intended issues was not fully realized. This limitation should be considered when interpreting the findings, as further work is needed to complete the analysis of the remaining data quality issues.
\begin{table*}[ht]
\centering
\caption{Iterations and Errors for Different Visualizations of Phase 1}
\label{tab:visualizations_errors}
\begin{tabular}{|c|l|c|c|l|}
\hline
\textbf{No.} & \textbf{Visualization Type}  & \textbf{Iterations} & \textbf{Errors} & \textbf{Error Type} \\ \hline
1  & Line Graph                & 1 & 0 & 0        \\ \hline
2  & Heat Map                  & 1 & 0 & 0        \\ \hline
3  & Bar Chart - Frequency      & 2 & 1 & KeyError \\ \hline
4  & Map                       & 2 & 1 & Runtime  \\ \hline
5  & Heat Map                  & 1 & 0 & 0        \\ \hline
6  & Bar Chart - Comparison     & 2 & 1 & Visual   \\ \hline
7  & Word Cloud                & 2 & 1 & Visual   \\ \hline
8  & Pie Chart                 & 2 & 1 & Visual   \\ \hline
9  & Word Cloud                & 1 & 0 & 0        \\ \hline
10 & Line Graph                & 2 & 1 & TypeError\\ \hline
\end{tabular}
\end{table*}

\subsection{Data Analysis}
\subsubsection{Approach and Framework}

The analysis focused on understanding the effects of data quality issues on AI-assisted data visualizations and their impact on users who rely on these visualizations for decision-making. The study adopted a user-centered approach, wherein each visualization was evaluated from the perspective of a user encountering these data issues. This involved both qualitative and observational analysis to assess the clarity, accuracy, and reliability of the visualizations produced.
\subsubsection{Evaluation Criteria}

The analysis was guided by the following key criteria:
\begin{itemize}
    \item Visual Integrity: Assessing whether the visualization accurately represents the data. This includes checking for distortions, misleading representations, or inaccuracies caused by data quality issues.
    \item Clarity and Comprehensibility: Evaluating how easily a user can understand the visualization. This involves considering whether the data quality issues resulted in visual clutter, confusing graphics, or an overwhelming amount of information.
    \item Insightfulness: Determining the extent to which the visualization provides meaningful insights. This criterion assesses whether the data quality issues obscure or highlight trends and patterns in ways that might mislead or confuse a user.
    \item User Confidence: Assessing how the presence of data quality issues might affect a user’s trust and confidence in the visualization. This involves considering whether a user would find the visualization reliable or useful for decision-making, given the observed data issues.
\end{itemize}

\subsubsection{Analytical Procedure}
\begin{enumerate}
    \item Observation of Effects: For each visualization created, detailed observations were recorded regarding how specific data quality issues (such as missing data, duplicate data, or inconsistent data types) impacted the visual output. This included noting any anomalies, unexpected results, or misrepresentations that occurred due to these issues.
    \item Impact Assessment: Placing oneself in the role of a user, the analysis focused on the potential cognitive and interpretative challenges these visualizations would present. This involved reflecting on how a user might perceive and react to the inaccuracies or complications caused by the data quality problems.
    \item Categorization of Issues: The observed effects were categorized based on the type of data quality issue and the nature of the visualization error or distortion it caused. This categorization helped in identifying common patterns and understanding the specific ways in which different data issues affect visualization outcomes.
    \item Iterative Feedback Loop: Using an iterative approach, each visualization was reassessed after corrections were suggested and implemented by ChatGPT. This helped in evaluating the effectiveness of the AI in addressing the identified data issues and understanding the limitations of AI-assisted visualizations in handling flawed data

\end{enumerate}

\section{Results}

This section presents the findings of the study, organized into three phases: analysis of a clean dataset, analysis of an unclean dataset, and an experimental study with injected data quality issues.

\subsection{Phase 1: Clean Dataset Analysis Results}
Visualizations created from the clean dataset largely adhered to the expected results, with only minor corrections needed in some cases. Table~\ref{tab:visualizations_errors} shows the types of visualizations created along with some metrics.

The visualizations created in this phase were mostly accurate, though a few errors were encountered during the generation process, which were either corrected by ChatGPT or flagged by the user. These errors can be classified into two categories: code errors and visual errors.
The first category, code errors, included errors such as KeyError, RuntimeError, and TypeError, which were immediately identified and thrown by the Python script. For instance, in the bar chart (visualization \#3), a KeyError occurred due to incorrect column referencing, which was automatically corrected in the second iteration after the code was adjusted. Similarly, in the map visualization (\#4), a RuntimeError was encountered because the mapping library failed to load due to missing data references, which was resolved in the second attempt by fixing the data input format. These types of errors were easily detectable by the system, allowing ChatGPT to assist in correcting them after the initial iteration.

The second category, visual errors, were not caught by the code itself but were noticeable as a user in the final visualization output. In these cases, the Python code executed successfully without error messages, but the resulting visualizations were flawed or unreadable due to incorrect data interpretation. For example, in the comparison bar chart (visualization \#6), too many bars were generated, resulting in an overcrowded chart, which made it unreadable. This issue occurred because the dataset was incorrectly interpreted, causing the visualization to include extraneous categories that were not relevant. This error required user intervention to notice and correct.

In visualization \#7 (word cloud), the type of data being displayed was incorrect—words that should not have appeared in the word cloud were included due to a mismatch in the dataset columns used for analysis. This issue was not flagged by the code, but the user had to manually identify the error and modify the dataset input for the visualization to reflect the correct words.
Additionally, in visualization \#8 (pie chart), the zip codes were incorrectly displayed as decimal numbers instead of integers. Although the visualization was technically correct in terms of structure, this formatting issue can cause different problems for a user when interpreting it. In this case, the error did not affect the logic of the code, but it created a misleading or unclear visualization that had to be manually corrected.

Overall, the errors in this phase were either detected automatically by the code (and subsequently fixed by ChatGPT) or noticed by the user in the visual outputs, requiring manual corrections. The visual errors, in particular, highlight the importance of user oversight in ensuring that AI-assisted visualizations accurately reflect the intended data.

\begin{table*}[ht]
\centering
\caption{Analysis Summary with Difficulty Scale (1 is least difficult and 5 is most difficult)}
\label{tab:analysis}
\begin{tabular}{|c|l|l|c|c|c|}
\hline
\textbf{No.} & \textbf{Analysis}                              & \textbf{Type}        & \textbf{Iterations} & \textbf{Time (mins)} & \textbf{Difficulty (1-5)} \\ \hline
1            & Artwork Distribution by Department             & Bar Chart            & 2                   & 10                   & 1                         \\ \hline
2            & Temporal Distribution of Artworks              & Histogram            & 5                   & 60                   & 3                         \\ \hline
3            & Culture Composition                            & Pie Chart            & 6                   & 180                  & 5                         \\ \hline
4            & Geographical Representation                    & Heat Map             & 4                   & 300                  & 4                         \\ \hline
5            & Text Analysis of Medium                        & Word Cloud           & 5                   & 60                   & 2                         \\ \hline
6            & Acquisition Analysis                           & Violin Plot          & 2                   & 30                   & 1                         \\ \hline
\end{tabular}
\end{table*}

\begin{table*}[ht]
\centering
\caption{Data Mirage Types and Visualization Errors}
\label{tab:mirages_errors}
\begin{tabular}{|c|l|l|}
\hline
\textbf{Vis.} & \textbf{Data Mirage Type(s)}                                                    & \textbf{Visualization Error}                                 \\ \hline
1                      & N/A                                                                            & Code Error                                                   \\ \hline
2                      & Inconsistent data types, missing data, poor aspect ratio                       & Caused data clumping in visual                               \\ \hline
3                      & Missing data, incorrect data formats, duplicate data                           & Data repetition in the chart, inaccurate portrayal of data    \\ \hline
4                      & Data inconsistencies, duplicated data, missing/null data                       & Overplotting                                                 \\ \hline
5                      & Duplicate data, irrelevant data                                                & Data repetition                                              \\ \hline
6                      & Missing data                                                                  & TypeError encountered                                        \\ \hline
\end{tabular}
\end{table*}

\subsection{Phase 2: Unclean Dataset Analysis Results}
In this phase, visualizations were generated using an unclean dataset from The Metropolitan Museum of Art Open Access dataset. Several data quality issues were present in this dataset, such as missing values, inconsistent data types, duplicate data, and incomplete documentation. These issues led to a variety of data mirages that distorted the accuracy and reliability of the visualizations. The following table summarizes the types of analyses performed, the iterations required to correct errors, and the difficulty of resolving each issue.

The visualizations created during this phase encountered several types of errors, both from code-related issues and visual inaccuracies caused by the underlying data. Data mirages, in particular, significantly distorted the interpretation of the visualizations. Below is a detailed summary of the specific data mirages, visualization errors, and the types of issues encountered from Table~\ref{tab:analysis}. Table\ref{tab:mirages_errors} breaks down the type of errors and mirages encountered at each visualization. 

\subsubsection{Overview of Errors and Data Mirages} 

The visualizations created during this phase encountered several types of errors, both from code-related issues and visual inaccuracies caused by the underlying data. Data mirages, in particular, significantly distorted the interpretation of the visualizations. Below is a detailed summary of the specific data mirages, visualization errors, and the types of issues encountered.

\subsubsection{Analysis:}
\begin{enumerate}
    \item \textbf{Artwork Distribution by Department (Bar Chart)}
    \begin{itemize}
        \item \textbf{Prompt:} “Create a bar chart using this dataset and information I gave you to visualize the distribution of artworks across different departments of the museum.”
        \item \textbf{Initial Code:} The initial code was provided by ChatGPT based on the prompt, which was designed to visualize the distribution of artworks across different museum departments.
        \item \textbf{Error Encountered:} A NameError was thrown when running the code, indicating that a variable or function had not been correctly defined.
        \item \textbf{Fix and Resolution:} ChatGPT was able to identify and correct the NameError in the second iteration. After making the necessary adjustments to the code (correcting the undefined variable), it successfully generated the desired bar chart.
        \item \textbf{Code After Fix:} The adjusted code (after the error was fixed) ran without issues and produced the correct visualization.
        \item \textbf{Time Taken:} 10 minutes.
        \item \textbf{Iterations}: 2 iterations were needed to resolve the issue and generate a proper bar chart fully.
        \item \textbf{Observation:} The error encountered was a basic coding mistake, which was quickly resolved by ChatGPT. Once fixed, the visualization was generated without any further complications. This particular visualization did not exhibit any data mirages, as the dataset was relatively clean for this analysis.
    \end{itemize}
    \item \textbf{Temporal Distribution of Artworks (Histogram)}
    \begin{itemize}
        \item \textbf{Prompt:} “Create a histogram showing the distribution of artworks by their object begin dates.”
        \item \textbf{Initial Code and Observations:} The initial code provided by ChatGPT resulted in an incorrect visualization, with all artworks grouped under the year “0” on the x-axis. This was due to inconsistencies in the “Object Begin Date” column of the dataset. The data contained various formats, such as numeric values, date ranges, and approximations (e.g., “1850,” “cr. 1850,” “1850-80,” and “ca. 1850”). These inconsistencies made it difficult for ChatGPT to generate an accurate histogram on the first attempt.
        \item \textbf{Error Encountered:} The visualization did not parse the object begin dates correctly, leading to the incorrect placement of all artworks under “0.”
        \item \textbf{Steps taken:}
        \begin{itemize}
            \item \textit{Feedback 1:} After observing the issue, I asked ChatGPT why all artworks were being placed under “0.” ChatGPT explained that the issue stemmed from how the dates were being read or processed.
            \item \textit{Feedback 2:} I informed ChatGPT that the dataset contained inconsistent date formats, such as “1850-80” and “cr. 1850-80,” and asked for an update to the code to account for these variations.
            \item \textit{Result:} ChatGPT updated the code but encountered a TypeError related to data types, which it successfully corrected in the next iteration.
            \item \textit{Feedback 3:} Despite this correction, the histogram still displayed incorrect data distribution, prompting further investigation of the dataset. I noticed additional date inconsistencies, such as “1894 ca.,” “1800,” “1730–40,” and “1800–1830 ca.” I informed ChatGPT of these new cases to handle a broader range of date formats.
            \item \textit{Solution Attempt:} ChatGPT proposed a more robust data parsing approach to address the diverse date formats. However, the final visualization still placed all data points under “0” due to improper x-axis formatting.
            \item \textit{Final Feedback:} I identified that the x-axis was not limited or formatted correctly to reflect the full date range in the dataset. After providing this feedback, ChatGPT adjusted the x-axis, and the correct distribution of artworks by their object begin dates was finally achieved.
        \end{itemize}
        \item \textbf{Time taken:}  AI-assisted time: 20-25 minutes with multiple iterations. Human intervention: Approximately 1-1.5 hours of manual investigation and data parsing.
        \item \textbf{Iterations:} 5
        \item \textbf{Level of Difficulty:} Moderate (required several rounds of feedback and human analysis).
        \item \textbf{Final Observations: } The main challenge stemmed from the inconsistent date formats in the “Object Begin Date” column. ChatGPT struggled to parse the data correctly, particularly when dates were not single integers but ranges or approximations. While the AI-generated code was functionally correct, the dataset’s complexity necessitated human intervention to identify and clean the data. The key takeaway from this task was that ChatGPT required more explicit information about the entire dataset (beyond the first 5 rows) to handle the inconsistencies in later rows. Human oversight was essential in examining the dataset and ensuring that all cases were accounted for before producing an accurate visualization.
    \end{itemize}
    \item \textbf{Culture Composition (Pie Chart)}
    \begin{itemize}
        \item \textbf{Prompt:} “Create a pie chart to display the composition of artworks by Culture (e.g., American, Mexican, etc.).”
        \item \textbf{Initial Code and Observations:} ChatGPT initially generated a pie chart based on the provided dataset, which appeared correct at first glance. However, upon closer examination of the "Culture" column, it became clear that the data contained multiple inconsistencies. These included entries with multiple cultures combined (e.g., “American or French”), entries with subcultures (e.g., “Chinese, for American market”), and entries with null values. Additionally, some entries had qualifying remarks (e.g., “French, probably”), making it difficult to group cultures accurately.
        \item \textbf{Challenges Encountered:}
        \begin{itemize}
            \item Data Inconsistencies: The dataset contained inconsistent entries, such as cultures combined into a single entry, subcultures listed after commas, and ambiguous values like “Attic” (a subcategory of Greek culture), which led to inaccurate results in the visualization.
            \item Missing Data: The dataset included many null cells, which resulted in missing data being ignored by the original visualization. An “unknown” category was created for these missing entries after further prompting.
            \item ValueError: During one of the iterations, a ValueError was thrown due to duplicate values in the dataset. This was resolved by resetting the index of the DataFrame to eliminate duplicates.
        \end{itemize}
        \item \textbf{Steps taken:}
        \begin{itemize}
            \item \textit{Initial Visualization:} A pie chart was generated, but the inconsistent and missing data led to inaccurate cultural representations. For example, subcultures such as “Attic” (Greek civilization) were listed as separate cultures rather than grouped with the broader "Greek" culture.
            \item \textit{Iteration 2:} After noticing the null data cells, I informed ChatGPT, which then created an “unknown” category for missing data. During this iteration, a ValueError was thrown and subsequently fixed by resetting the DataFrame’s index.
            \item \textit{Iteration 3:} I informed ChatGPT of the presence of entries like “American or French” and asked it to account for these types of cases in the visualization. ChatGPT provided an updated visualization, but further inconsistencies remained.
            \item \textit{Iteration 4:} While parsing the dataset, I discovered more complex entries, such as subcultures within general cultures (e.g., “Attic” and “Greek” listed separately). I asked ChatGPT to address this by grouping subcultures with their parent cultures. However, this required a deeper understanding of the dataset and cultural history, which was not fully handled by the AI.
            \item \textit{Iteration 5:} I pointed out additional inconsistencies, such as cultures with qualifiers and subcategories. ChatGPT made progress by creating grouped categories for these subcultures, but more complexities remained.
            \item \textit{Iteration 6:} Finally, I identified redundant entries, such as “Japan” and “Japanese” listed as separate categories, even though they referred to the same culture. I prompted ChatGPT to fix this, and a more accurate pie chart was produced.
        \end{itemize}
        \item \textbf{Time taken:} AI-assisted time: Approximately 3 hours, spread across 6 iterations. Human intervention: Significant manual input was required to identify and flag the numerous data inconsistencies.
        \item \textbf{Iterations:} 6 iterations were required to reach an acceptable solution.
        \item \textbf{Level of Difficulty:} Difficult (due to the complexity of data cleaning and categorization).
        \item \textbf{Final Observations:} The dataset presented significant challenges in handling subcultures and combined cultures. ChatGPT was able to handle some cases by creating mapping rules and managing a mapping dictionary to group similar cultures together (e.g., “Attic” and “Greek”). However, human oversight was essential for identifying new cases of repetition and inconsistency that did not appear in the first rows of the dataset. Even after multiple iterations, certain subcultures were still misrepresented in the visualization, suggesting that additional cases in the dataset were not covered. The process of cleaning the data and ensuring accurate cultural representation was time-consuming, highlighting the limitations of AI tools when working with complex, unclean datasets. Human intervention was necessary to review and manually inspect the entire dataset for these issues.

    \end{itemize}
    \item \textbf{Geographical Representation of Artworks (Heatmap)}
    \begin{itemize}
        \item \textbf{Prompt:} “Create a heatmap of the different cities/countries that each artwork came from.”
        \item \textbf{Data Observations:} The dataset contains several inconsistencies, including missing data and duplicate entries (e.g., "United States | England" or "United States | United States"). Additionally, there were entries where countries were uncertain or described with qualifiers (e.g., "France or North Spain," "possibly Syria," and "Southern Italy"), which made it difficult to correctly plot the origins of the artworks on the heatmap. Furthermore, differences between the dataset’s country names and those in the geopandas library added to the complexity of the task.
        \item \textbf{Iterations:}
        \begin{itemize}
            \item \textit{Iteration 1:} The initial heatmap was generated, but many countries had missing data. The dataset’s inconsistencies caused certain artworks to be incorrectly mapped or excluded from the heatmap.
            \item \textit{Iteration 2:} Upon reviewing the data further, I realized that many artworks were misrepresented due to the input format. For example, entries like "France or North Spain" and "Gaul (Northern France)" were not parsed properly, leading to an incorrect representation of the countries where the artworks originated. This resulted in a disproportionate number of artworks appearing to come from Egypt, while other countries were underrepresented.
            \item \textit{Iteration 3:} I asked ChatGPT why certain countries were missing on the map despite being present in the dataset. ChatGPT identified that this issue could be caused by discrepancies between the country names in the dataset and those in the geopandas library. For example, in the dataset, the USA is listed as "United States," while geopandas uses "United States of America." Older country names or inconsistent spellings also contributed to missing data on the map.
            \item \textit{Iteration 4:} After identifying and mapping some of the differences between the MET dataset and the geopandas dataset, ChatGPT was able to correct certain countries. However, many regions and civilizations were still not properly mapped. This was particularly evident for historical regions or subregions that did not match modern country boundaries in the geopandas library, such as "Byzantine Egypt" or "North Africa (box) | Spain (lid)."
        \end{itemize}
        \item \textbf{Time taken}: AI-assisted time: 5 hours. Human intervention: 2 days were required to manually identify and match the differences between country names in the dataset and the geopandas library.
        \item \textbf{Iterations:} 4 iterations were required to fix the major inconsistencies, though many issues remained unresolved due to the complexity of mapping historical regions and inconsistent data entries.
        \item \textbf{Level of Difficulty}: Moderately difficult (due to the need to match country names between two datasets and handle geographic inaccuracies).
        \item \textbf{Final Observations:} The main challenge in this visualization was aligning the dataset’s country names with those recognized by the geopandas library. For example, "United States" was not recognized by geopandas, causing the USA to be omitted from the map until it was corrected to "United States of America." Additionally, many historical or ambiguous regions (e.g., "Gaul," "possibly Syria") were not accounted for in the library, leading to missing data in the final visualization. Despite four iterations, many countries were still incorrectly mapped or missing. Human intervention was required to manually inspect the data and create mappings for each inconsistency. However, this task proved to be time-consuming and labor-intensive, as it involved reviewing the entire dataset to catch all the differences between the MET dataset and the geopandas country library. Although ChatGPT was able to assist in fixing some issues, resolving all the inconsistencies between the datasets required significant human input. Mapping historical or ambiguous regions was particularly challenging, as many of these regions do not have clear modern-day equivalents in the geopandas library. This task would likely require more comprehensive manual data cleaning and mapping to achieve a fully accurate heatmap.

    \end{itemize}
    \item \textbf{Text Analysis of Medium (Word Cloud)}
    \begin{itemize}
        \item \textbf{Prompt:} “Generate a word cloud visualization from the description of the mediums to show common materials and techniques from the 'Medium' column.”
        \item \textbf{Initial Code and Observations:} ChatGPT generated an initial word cloud based on the data from the "Medium" column of the dataset. While the word cloud appeared visually correct, a closer look revealed that several terms were duplicated (e.g., "silver," "terracotta," "commercial," "color," and "gelatin"). This duplication distorted the representation of the most common materials, as the same word appeared multiple times in the visualization.
        \item \textbf{Data Mirage:} The word cloud gave the false impression that certain terms were more frequent than they actually were, due to the repetition of the same words. This data mirage led to an inaccurate understanding of the frequency of materials and techniques used in the artworks.
        \item \textbf{Iterations:}
        \begin{itemize}
            \item \textit{Iteration 1:} The initial word cloud showed duplicated terms, which led to misrepresentations of the data.
            \item \textit{Iteration 2:} After prompting ChatGPT to remove the duplicates, some terms were corrected, but others remained. The issue of duplicated words persisted, with terms like "silver" and "color" still appearing multiple times.
            \item \textit{Iteration 3:} To address the persistent issue, ChatGPT implemented a tokenizer and stemmer to break down the words into root forms and attempt to eliminate duplicates. However, the duplicated terms continued to appear, suggesting that the underlying data may have had inconsistencies or variations in how the terms were stored (e.g., case sensitivity or minor spelling differences).
            \item \textit{Iteration 4:} In this final iteration, I prompted ChatGPT to clean up irrelevant terms and further refine the word cloud by ensuring that only unique, meaningful terms were displayed. The result was a more accurate word cloud, but some minor issues with term duplication remained due to the complexity of handling variations in the dataset.
        \end{itemize}
        \item \textbf{Time Taken:} AI-assisted time: 60 minutes. Human Intervention: 3 hours
        \item \textbf{Iterations:} 4 iterations were required to reach a satisfactory visualization.
        \item \textbf{Level of Difficulty:} Moderate (due to the challenge of removing duplicated terms and cleaning irrelevant words).
        \item \textbf{Final Observations:} The primary challenge in this visualization was the removal of duplicated terms that were appearing due to variations in the dataset, such as differences in capitalization or minor spelling differences (e.g., “silver” and “Silver”). ChatGPT’s implementation of tokenization and stemming partially resolved the issue but was not able to completely eliminate all duplicate terms. Additionally, irrelevant terms (such as prepositions or articles) had to be manually removed to ensure that only meaningful materials and techniques were represented in the word cloud. While the final word cloud was much cleaner, the presence of minor duplications indicates that further data preprocessing might be required for larger or more complex datasets to ensure full accuracy.
    \end{itemize}
    \item \textbf{Acquisition Analysis (Violin Plot)}
    \begin{itemize}
        \item \textbf{Prompt:} “Create a violin plot to visualize the distribution of acquisition methods used over time for this dataset.”
        \item \textbf{Data Insights:} The AccessionYear column, which contains the dates of acquisition, is mostly consistent, though there are some null data cells. The Credit Line column, which describes the acquisition method, is entered as descriptive text (e.g., “Gift of Mr. and Mrs. John Doe”). ChatGPT’s initial approach involved splitting the acquisition methods into three categories: gift, purchase, and other. This basic classification does not account for the full complexity of the descriptions in the Credit Line column.
        \item \textbf{Iterations:}
        \begin{itemize}
            \item \textit{First Iteration:} The initial code provided by ChatGPT threw a TypeError. This error was likely due to issues handling the null values in the AccessionYear column or misinterpreting certain entries in the Credit Line column.
            \item \textit{Second Iteration:} After prompting ChatGPT to address the error, the TypeError was resolved, and a violin plot was successfully generated. The plot displayed the distribution of acquisition methods (categorized into gift, purchase, and other) over time.
        \end{itemize}
        \item \textbf{Time taken:} AI-assisted time: 30 minutes (including error correction and generation).
        \item \textbf{Iterations:} 2 iterations were required to fix the error and generate the correct visualization.
        \item \textbf{Level of Difficulty:} Easy (as the errors were relatively simple to resolve, and the dataset was mostly clean in terms of the AccessionYear column).
        \item \textbf{Final Observations:} 
        While the violin plot was successfully generated after the second iteration, the categorization of acquisition methods was too simplified. ChatGPT only distinguished between gift, purchase, and other, which does not capture the full range of acquisition methods described in the Credit Line column (e.g., donations, bequests, transfers). A more detailed classification of acquisition methods could provide additional insights into the distribution. The presence of null data in the AccessionYear column did not significantly impact the visualization after the error was fixed, though more thorough handling of missing data could improve the robustness of the plot. Overall, the visualization was produced without major issues, but there is room for improvement in how acquisition methods are categorized and represented.
    \end{itemize}
\end{enumerate}

\subsection{Phase 3: Experimental Study with Injected Data Quality Issues Results}
In this phase, specific data quality issues were systematically injected into ten clean datasets. The aim was to observe how these issues impacted the AI-generated visualizations and evaluate the AI’s ability to handle them. The following eight data quality issues were introduced into the datasets:

\begin{itemize}
    \item Missing data
    \item Duplicate data
    \item Inconsistent data  
    \item Inconsistent data types
    \item Inaccurate data
    \item Irrelevant data
    \item Data entry errors
    \item Incorrect data formats

\end{itemize}
Each dataset was used to generate five different types of visualizations: pie charts, word clouds, histograms, heat maps, and scatter plots. The results show how these data quality issues impacted the visualization outcomes and whether AI was able to correct them.
\\

\textbf{Dataset 1: Most Streamed Spotify Songs 2024}
\\
The first dataset used in this phase was the "Most Streamed Spotify Songs 2024" dataset from Kaggle.
Results from the clean dataset:
\begin{itemize}
    \item Visualization 1: Bar Graph: The bar graph was generated successfully, representing the top 10 most streamed artists on Spotify.
    \item Visualization 2: Line Graph: The line graph displayed the streaming trends of the top artists over time without issue.
    \item Visualization 3: Word Cloud: The word cloud accurately represented the most common artist names based on the dataset.
    \item Visualization 4: Heat Map: The heat map successfully visualized the geographical distribution of streams for the top artists.
    \item Visualization 5: Pie Chart: The pie chart showed the distribution of streaming numbers for the top artists.

\end{itemize}
\subsubsection{Data Quality Issue 1: Missing Data}

 The experiment simulated missing data by removing 15\% of the dataset at random using a Python script. Prior to injecting the data quality issue, the five visualizations were run on the clean dataset as a baseline for comparison.
Encoding Issue: An initial encoding issue was encountered when reading the dataset, resulting in incorrect characters appearing in certain visualizations. This was resolved by specifying an encoding format when reading the dataset, and by applying a cleaning script to remove unwanted characters from the "Artist" column.

After injecting 15\% missing data into the dataset, the five visualizations were re-run to observe the impact of the missing values.
\begin{enumerate}
    \item \textbf{Visualization 1: Bar Graph:}  The bar graph remained largely unaffected by the missing data, with the top 10 artists displaying consistently with only minor variations.
    \item \textbf{Visualization 2:Line Graph:}  The line graph was similarly unaffected, as the overall streaming trends did not change significantly with the missing data.
    \item \textbf{Visualization 3: Word Cloud:}  The word cloud was noticeably impacted by the missing data. Several artist names appeared more frequently due to the removal of data for other artists.  Additionally, encoding errors reappeared, resulting in the display of unwanted characters in place of certain artist names. After cleaning, the word cloud showed a skewed distribution of artist names compared to the clean dataset, due to missing data.
    \item \textbf{Visualization 4: Heat Map:} The heat map was impacted by the missing data. Several artists had different streaming values, which altered the overall distribution of streams. The top 10 artists displayed in the heat map changed when compared to the clean dataset.
    \item \textbf{Visualization 5: Pie Chart:} The pie chart displayed altered proportions after the missing data was injected, causing changes in the relative ranking of artists. The overall order of artists in the pie chart differed from the clean dataset, highlighting the effect of missing data on the representation of streaming numbers.
\end{enumerate}

\subsubsection{Data Quality Issue 2: Duplicate Data
}
To simulate the effects of duplicate data on visualizations, a new dataset was created by injecting random duplicated rows into the clean Spotify dataset. This duplication was designed to observe how the visualizations would be affected by redundant information.

The five visualizations were re-run with the duplicated dataset, and the following differences were observed compared to the clean dataset:

\begin{enumerate}
    \item \textbf{Visualization 1: Bar Chart:} The bar chart was affected by the duplicated data, with fewer tracks appearing in the final visualization. This is because several songs with duplicate data were tied with the same number of streams, causing them to compete for representation in the chart, thus removing some tracks entirely.
    \item \textbf{Visualization 2: Line Graph:} At first glance, the line graph appeared similar between the clean and duplicated datasets. However, upon closer inspection, there were subtle differences in the points representing streaming trends over time. These small differences likely arose due to the duplication affecting certain song data, but the overall trends remained consistent.
    \item \textbf{Visualization 3: Word Cloud}: The word cloud showed a clear impact from the duplication of data. Some artists appeared more frequently than in the clean dataset, giving the false impression that they were more prevalent. This duplication caused other artists to be underrepresented compared to the original dataset, creating an imbalance in the visual representation.
    \item \textbf{Visualization 4: Heat Map:}  The heat map, which visualized the top 10 artists versus various metrics (e.g., streams, albums, popularity), was affected by the duplicated data. The duplication of rows resulted in certain artists being over-represented in the heat map, with inflated values for metrics such as total streams. As a result, some artists appeared to perform better than they actually did, while others were underrepresented due to the duplicates skewing the distribution.
    \item \textbf{Visualization 5: Pie Chart:} The pie chart displayed noticeable shifts in the percentages associated with various artists. Due to the duplication, the streaming percentages for some artists increased, while others decreased, altering the overall ranking of the artists in the pie chart.
\end{enumerate}

\subsubsection{Data Quality Issue 3: Inconsistent Data}

To simulate the effect of inconsistent data, a script was used to modify the artist names in the dataset by changing the capitalization across 15\% of the data and removing letters to create inconsistent spelling. This was intended to observe how these inconsistencies impacted the visualizations and whether they would lead to repeated entries or inaccurate representations of the data. The five visualizations were re-run with the dataset containing inconsistent data, and the following differences were observed compared to the clean dataset:

\begin{enumerate}
    \item \textbf{Visualization 1: Bar Graph:}
    The bar graph was affected by the inconsistent artist names, leading to certain songs being listed twice. For example, “As It Was” by Harry Styles appeared twice due to the inconsistent capitalization of the artist’s name. This duplication reduced the available space for other entries in the visualization and led to inaccurate representations of the top tracks. Impact: This duplication could cause users to miss valuable insights by presenting redundant information, which could affect decision-making based on the visualization.
    \item \textbf{Visualization 2: Line Graph:} The line graph was unaffected by the inconsistent data because the graph used parameters unrelated to the artist's names, such as streaming counts over time. Impact: No significant effect was observed due to inconsistent data in this case.
    \item \textbf{Visualization 3: Word Cloud:} Despite the initial hypothesis that the word cloud would be heavily impacted by inconsistent data, the visualization remained largely unaffected. This could be due to the word cloud library’s built-in functionality that includes stop words and filters that may automatically normalize or exclude minor inconsistencies in the data. Impact: The word cloud did not show noticeable changes between the clean and inconsistent datasets, suggesting that this type of visualization might be more robust against small inconsistencies in text data.
    \item \textbf{Visualization 4: Heat Map:} The heat map, which visualized the top 10 artists versus various metrics (e.g., streams, album releases), was significantly affected by the inconsistent data. Inconsistencies in artist names resulted in certain artists being duplicated or omitted entirely from the heat map. For example, an artist like "Drake" might have been split into two entries: "Drake" and "drake," which skewed the visual representation of the top artists. Impact: This led to a misrepresentation of the top 10 artists, where certain artists appeared multiple times under different names, while others were underrepresented or omitted. This could result in an inaccurate analysis of artist performance across metrics and mislead users into drawing incorrect conclusions about the most successful artists.
    \item \textbf{Visualization 5: Pie Chart:} At first glance, the pie charts from the clean and inconsistent datasets appeared similar. However, closer inspection revealed that the inconsistent artist names caused minor shifts in the percentages and rankings of the top 10 artists. While these differences were subtle, they introduced new artists into the top 10, which altered the overall distribution slightly. Impact: Though the changes were small, the inconsistency could affect the user’s interpretation of the data by misrepresenting which artists were the most streamed. Over time, this could skew the analysis, especially in datasets where accuracy is critical.
\end{enumerate}

\subsubsection{Data Quality Issue: Inconsistent Data Types}
For this experiment, inconsistent data types were simulated by mixing numerical data (numerals) and string values in the dataset. This introduced conflicting formats in numerical columns, particularly affecting those that rely on consistent numerical data (e.g., streams, ranking positions). The goal was to assess how inconsistent data types impact the generation of different visualizations and whether AI-assisted tools could handle the mixed data types.

Results After Injecting Inconsistent Data Types:

\begin{enumerate}
    \item \textbf{Visualization 1: Bar Graph:} The bar graph failed to generate due to the presence of both strings and numerals in the "Spotify Streams" column. Upon the first attempt, the graph returned an empty visualization. In the second iteration, an error message was displayed: “No valid numeric data available in 'Spotify Streams' to display.” This highlights the challenge in producing the visualization when data types are inconsistent, rendering the bar graph unusable. Impact: The inconsistency in data types prevents the user from obtaining a bar graph for analysis, which would require manual correction of the data format before proceeding with any meaningful visualizations.
    \item \textbf{Visualization 2: Line Graph:} Similar to the bar graph, the line graph also failed to generate due to the mixed data types. An error was thrown during the first iteration, preventing the visualization from being displayed. Impact: The inconsistent data types in the dataset prevented the generation of the line graph, and no meaningful visualization was produced.
    \item \textbf{Visualization 3: Word Cloud:} Despite the numerical data inconsistencies, the word cloud was generated. The impact of the inconsistent data types was minimal, with only slight changes observed in the size of certain artists' names. For example, "Ariana Grande" and "Billie Eilish" appeared slightly larger compared to the clean dataset, but the difference was not very noticeable. Impact: The word cloud proved more robust against the mixed data types. However, slight changes in the size of artist names indicated that the inconsistency did have a minor impact on the representation of the data.
    \item \textbf{Visualization 4: Heat Map:} The heat map failed to generate due to the inconsistent data types. An error was encountered, which prevented the visualization from being displayed. Impact: The mixed formats in the dataset caused the heat map to fail, as the data required for the visualization could not be processed due to the conflicting data types.
    \item \textbf{Visualization 5: Pie Chart:} Similar to the heat map, the pie chart also failed to generate. The same error related to the inconsistent data types prevented the pie chart from displaying any results. Impact: The pie chart was rendered unusable due to the mixed data formats in the numerical column, requiring the user to clean the data before being able to visualize the correct results.
\end{enumerate}
Inconsistent data types had a significant impact on most of the visualizations. The bar graph, line graph, heat map, and pie chart all failed to generate due to the mixed formats in the "Spotify Streams" column, displaying error messages or returning empty results. The word cloud was the only visualization that managed to display some results, though with minimal changes. These findings demonstrate that inconsistent data types can render certain visualizations completely unusable and highlight the importance of consistent formatting in numerical datasets when generating accurate visualizations.

\subsection{Summary of Findings:}
In total, the results demonstrated that AI-assisted tools like ChatGPT are useful for generating data visualizations, particularly when working with clean datasets. However, human intervention is crucial when handling unclean datasets or correcting more complex data quality issues. Across the five visualizations, the most common challenges arose from missing data, duplicated data, inconsistent data, and inconsistent data types, each affecting visualization accuracy in different ways.

While ChatGPT provided functional code for most tasks, additional steps were required to address deeper issues within the dataset that impacted the accuracy and reliability of the visualizations. Data quality issues such as inconsistent data types and duplicated entries led to visualization failures or misrepresentations, highlighting the limitations of AI tools in fully addressing these complexities on their own.

The results of this study reveal both the potential and limitations of using AI-assisted tools for data visualization in the presence of data quality issues. Although these tools can assist in generating visualizations, significant manual effort is needed to correct underlying data inconsistencies, ensuring accurate representation. In the following section, we will discuss these findings in greater detail, considering the broader implications for AI-driven visualization processes and areas where future improvements are necessary.

\section{Discussion}
After conducting each phase of the experiment to better understand how ChatGPT can assist in data visualization creation and how different data quality issues affect these visualizations, valuable insights were gained. This study aimed to explore the capabilities and limitations of AI-assisted tools in generating accurate visualizations when faced with real-world data quality challenges.

\subsection{Key Findings}
There was notable potential in the Python scripts generated by ChatGPT. In most cases, the code provided was technically correct, successfully producing visualizations from clean datasets. However, when various data quality issues were introduced—such as missing, duplicated, and inconsistent data—the visualizations were significantly affected. These edge cases highlight the challenge users face when relying solely on AI tools for visualization tasks without understanding or addressing underlying data issues.

The experiment positioned me in the role of a typical user who is unfamiliar with data quality issues and simply wants a visualization for analysis. This study bridges the gap between AI-generated visualizations and the potential roadblocks that amateur users might face when trying to generate meaningful insights from unclean data. Identifying these issues opens the door for future improvements in AI tools, making them more accessible and user-friendly for non-experts.

When confronted with data quality issues, the visualizations often misrepresented key data points, leading to false interpretations. For instance, missing data skewed pie charts and heatmaps, while duplicated data distorted bar charts and word clouds. Inconsistent data types prevented certain visualizations from being generated altogether. These findings suggest that while AI tools like ChatGPT can produce functional visualizations, they lack robust mechanisms to handle data quality issues, often requiring human intervention to correct errors and ensure accuracy.

\subsection{Comparison to Previous Research}
These results align with findings from previous studies of \cite{vizmirages}, which emphasized the difficulties in handling inconsistent data in AI-generated visualizations. However, this study goes further by exploring the specific effects of duplicated data, an area that has received less attention in the literature. While previous research focused primarily on inconsistency and bias, this study highlights how data duplication can equally distort AI-generated visualizations, particularly in metrics-based representations such as bar graphs and pie charts.

\subsection{Limitations of the Study}
A limitation of this study is the narrow scope of data quality issues examined, focusing only on missing data, duplicated data, and inconsistent data types. Other potential issues, such as data outliers, mislabeling, or incorrect categorizations, were not included. Addressing a wider range of data quality problems could have provided deeper insights into the limitations of AI in handling unclean data. Moreover, this study relied exclusively on ChatGPT, which may limit the generalization of the findings to other AI-assisted visualization platforms. Future studies could involve comparisons between different AI tools to see if similar challenges persist across platforms.

The 10-week time-frame also limited the extent of the analysis. Time constraints prevented the exploration of additional datasets and data quality issues, which could have enriched the study’s findings and provided further evidence of AI tools' capabilities and limitations in data visualization.

\subsection{Implications for AI-Assisted Data Visualization}
These findings have significant implications for industries that rely on AI-assisted data visualization tools. While AI can expedite the visualization process, this study reveals that the presence of unclean data often leads to inaccurate results, necessitating human oversight. The necessity of manual intervention to correct errors and ensure the accuracy of the visualizations highlights that AI tools are not yet capable of fully replacing human expertise in data analysis and visualization.

For non-expert users, unclean data poses a significant barrier to effective analysis using AI-generated visualizations. This study suggests that more work is needed to make AI tools capable of detecting and resolving common data quality issues autonomously.

\subsection{Future Research Directions}
Future research could focus on enhancing the capabilities of AI tools to detect and automatically correct unclean datasets before generating visualizations. This could involve integrating more advanced data cleaning and error detection algorithms into AI models. Additionally, expanding the scope of the research to include other types of data quality issues—such as data outliers, mislabeled data, and inconsistencies in time series—would provide a more comprehensive understanding of how AI tools can adapt to different data challenges.

Another promising direction for future research would be to compare different AI models to determine their effectiveness in handling data quality issues. By testing multiple tools in a variety of real-world contexts, researchers can better assess the potential of AI to fully automate the visualization process and reduce the need for human intervention.

\section{Conclusion}
This study set out to explore the capabilities and limitations of AI-assisted data visualization tools, focusing on how ChatGPT handles common data quality issues such as missing, duplicated, and inconsistent data. Through a series of experiments, it became clear that while AI tools are effective at generating visualizations from clean datasets, they struggle significantly when faced with unclean data. Human intervention remains necessary to ensure accuracy, especially when dealing with complex data issues.
The findings have important implications for non-professional users who rely on AI tools to visualize data. In particular, flawed visualizations caused by unclean data can lead to misinterpretations and inaccurate conclusions, which are particularly problematic for users without expertise in data quality management. When key insights are distorted due to issues like duplication or inconsistent data types, non-professional users may unknowingly base their analysis on flawed representations, potentially leading to poor decision-making or incorrect outcomes. This highlights a critical gap in AI-assisted visualization tools, where users must be aware of data quality issues and possess the ability to intervene.

Despite its limitations, this study contributes valuable insights into the growing field of AI-assisted data visualization. By identifying key challenges, it opens the door for future advancements in AI-driven tools that are better equipped to handle data quality issues autonomously, especially for users who lack technical expertise. Improvements in these tools, such as built-in error detection and automatic data cleaning, could greatly enhance the user experience by making data visualization more accessible and reliable.

For now, while AI tools have the potential to streamline the data visualization process, they remain complementary to human oversight. The need for manual data correction underscores the importance of human expertise in ensuring accurate and meaningful visual outputs. With further improvements, AI-assisted visualization tools could become indispensable, empowering users across a wide range of industries and expertise levels to generate accurate and insightful visualizations from their data.

\begin{acks}
Thank you to Dr. Anna Fariha for her guidance and mentorship through this project. This research would not have been possible without the support of the CRA (Computing Research Association) through the 2024 DREU program.
\end{acks}

\bibliographystyle{ACM-Reference-Format}
\bibliography{biblo}

\end{document}